\documentclass[12pt]{article}
\usepackage[latin9]{inputenc}
\usepackage[letterpaper]{geometry}
\usepackage{float}
\usepackage{amsmath}
\usepackage{amssymb}
\usepackage{graphicx}
\usepackage{esint}

\makeatletter
\newcommand{\lyxaddress}[1]{
\par {\raggedright #1
\vspace{1.4em}
\noindent\par}
}

\@ifundefined{date}{}{\date{}}
\usepackage{epstopdf}
\numberwithin{equation}{section}

\makeatother

\begin{document}

\title{$\eta-\eta'$ mixing and the derivative of the topological susceptibility
at zero momentum transfer}

\author{N. F. Nasrallah}
\maketitle

\lyxaddress{}

\lyxaddress{\begin{center}
Faculty of Science, Lebanese University. Tripoli 1300, Lebanon 
\par\end{center}}
\begin{abstract}
The couplings of the isosinglet axial-vector currents to the $\eta$
and $\eta'$ mesons are evaluated in a stable, model independent way
by use of polynomial kernels in dispersion integrals. The corrections
to the Gell-Mann-Oakes-Renner relation in the isoscalar channel are
deduced. The derivative of the topological susceptibility at the origin
is calculated taking into account instantons and instanton screening. 
\end{abstract}

\section{Introduction}

The subject of $\eta-\eta'$ mixing has been a topic of discussion
since $SU(3)$ flavor symmetry was proposed \cite{PhysRevLett.55.2766,ball1996phenomenological,akhoury1989eta,feldmann1998mixing,nasrallah2004glue,escrib}.
The gluon axial anomaly and the corresponding topological charges
of the isoscalor mesons imply that the $SU(3)$ singlet axial vector
current is not conserved in the chiral limit. Initially the octet-singlet
mixing was described by an angle $\theta$ which was thought to be
small and later given larger values \cite{PhysRevLett.55.2766}.It
was later realized that the couplings of the isoscalar axial currents
to the pseudoscalar mesons need not be dependent and that the single
angle description is inadequate. A number of theoretical approaches
have been used to compute these couplings. Apart from Chiral perturbation
theory \cite{kaiser2000large} QCD sum rules \cite{narison1999topological},
Shore \cite{shore2006pseudoscalar} has used the generalized Gell-Mann-Oakes-Renner
\cite{PhysRev.175.2195} relation to evaluate the couplings.

A related topic is the calculation of the topological susceptibility
and its derivative at zero momentum transfer. The results obtained
show a wide dispersion \cite{ioffe2000correlation,pasupathy2006derivative,singh2009axial}.
Such a dispersion in the results and instabilities in the parameters
which enter the calculations is inherent in the Borel (Laplace) sum
rules \cite{SHIFMAN1979385} used by the authors.

This method starts from a dispersion integral. 
\begin{equation}
Residue=\dfrac{1}{\pi}\int_{th}^{\infty}dt\ e^{-t/M^{2}}Im\ P(t)\label{eq1.1}
\end{equation}

The residue contains the physical quantity of interest and the integral
runs from the physical threshold to infinity. The integral is then
split into two parts 
\begin{equation}
\int_{th}^{\infty}dt\ e^{-t/M^{2}}Im\ P(t)=\int_{th}^{t_{0}}dt\ e^{-t/M^{2}}Im\ P(t)+\int_{t_{0}}^{\infty}dt\ e^{-t/M^{2}}Im\ P(t)\label{eq1.2}
\end{equation}
where $t_{0}$ signals the onset of perturbative QCD. In the first
integral on the r.h.s of the equation above $ImP(t)$ describes the
unknown contribution of the resonances. The second integral takes
into account the contribution of the QCD part of the amplitude when
$P(t)$ is replaced by its QCD expression. $M^{2}$, the square of
the Borel mass is a parameter introduced in order to suppress the
unknowns of the problem. If $M^{2}$ is small, the damping of the
first unknown integral is good but the contribution of the unknown
higher order non perturbative condensates increases rapidly. If $M^{2}$
increases, the contribution of the unknown condensates decreases but
the damping in the resonances region worsens. An intermediate value
of $M^{2}$ has to be chosen. Because $M^{2}$ is a non physical parameter
the results should be independent of it in a relatively broad window;
this is not the case in the problems at hand. The choice of the parameter
$t_{0}$ which signals the onset of perturbative QCD is another source
of uncertainty. In this work I shall use low order polynomial kernels
in order to suppress the contribution of the unknown continuum. The
coefficients of these polynomials are determined by the masses of
the isoscalar resonances and the method avoids the instabilities and
arbitrariness which accompany the use of exponential kernels. Having
determined the couplings of the isoscalar currents to the $\eta$
and $\eta'$ mesons I shall turn to the study of the corrections to
the Gell-Mann-Oakes-Renner relation \cite{PhysRev.175.2195} in the
isoscaler channel and recover $m_{\eta}$. Finally I shall evaluate
$\chi'(0)$ the derivative of the topological susceptibility at zero
momentum transfer taking into account the effect of instantons and
their possible screening which can be important as has been emphasized
by Forkel \cite{forkel2005direct}.

\section{Axial currents and their coupling to the $\eta-\eta'$ mesons}

The isoscalar components of the octet of axial vector currents couple
to the physical $\eta$ and $\eta'$ mesons: 
\begin{equation}
\begin{aligned}\langle0|A_{\mu}^{(8)}|\eta(p)\rangle= & 2if_{8\eta}p_{\mu}\\
\langle0|A_{\mu}^{(0)}|\eta(p)\rangle= & 2if_{0\eta}p_{\mu}\\
\langle0|A_{\mu}^{(8)}|\eta'(p)\rangle= & 2if_{8\eta'}p_{\mu}\\
\langle0|A_{\mu}^{(0)}|\eta'(p)\rangle= & 2if_{0\eta'}p_{\mu}
\end{aligned}
\label{eq2.1}
\end{equation}

In the $SU(3)$ limit $f_{8\eta}=f_{\pi}=92.4MeV$ and in the two
mixing angle description adopted here, the coupling constants above
are independent quantities. The axial vector currents are written
in terms of the quark fields :

\begin{equation}
\begin{aligned}A_{\mu}^{(8)}= & \frac{1}{\sqrt{3}}\ (\bar{u}\gamma_{\mu}\gamma_{5}u+\bar{d}\gamma_{\mu}\gamma_{5}d-2\bar{s}\gamma_{\mu}\gamma_{5}s)\\
A_{\mu}^{0}= & \sqrt{\frac{2}{3}}\ (\bar{u}\gamma_{\mu}\gamma_{5}u+\bar{d}\gamma_{\mu}\gamma_{5}d+\bar{s}\gamma_{\mu}\gamma_{5}s)
\end{aligned}
\label{eq2.2}
\end{equation}

In the limit $m_{u}=m_{d}=0$, the divergences of these currents are

\begin{equation}
\begin{aligned}\partial_{\mu}A_{\mu}^{8}= & \frac{2}{\sqrt{3}}(-2im_{s}\bar{s}\gamma_{5}s)\\
\partial_{\mu}A_{\mu}^{0}= & -\sqrt{\frac{2}{3}}(-2im_{s}\bar{s}\gamma_{5}s)+2\sqrt{6}Q
\end{aligned}
\label{eq2.3}
\end{equation}
where $Q=\frac{\alpha_{s}}{8\pi}G\widetilde{G.}$ is the anomaly with
$G\tilde{G}=G_{\mu\nu}\tilde{G}^{r\gamma}$, $G_{\mu\nu}$ being the
gluon field strength tensor and $\tilde{G}_{\mu\nu}=\frac{1}{2}\epsilon_{\mu\nu p\sigma}\ G^{p\sigma}$
its dual. Consider now the correlator:

\begin{equation}
\Pi_{\mu\nu}^{ij}=\int dx\ e^{iqx}\langle\ 0\ |\ T\ A_{\mu}^{(i)}(x)\ A_{\nu}^{(j)}\ (0)\ |\ 0\ \rangle\label{eq2.4}
\end{equation}
$i,j=0,8$

It can be decomposed 
\begin{equation}
\Pi_{\mu\nu}(q^{2})=(-g_{\mu\nu}q^{2}+q_{\mu}q_{\nu})\ \Pi(q^{2})^{(1)}+q_{\mu}q_{\nu}\ \Pi(q^{2})^{(0)}\label{eq2.5}
\end{equation}
and let 
\begin{equation}
\Pi(t=q^{2})=\Pi^{(1)}(t)+\Pi^{(0)}(t)\label{eq2.6}
\end{equation}

Start with $\Pi^{88}(t)$. At low energies it has two poles 
\begin{equation}
\Pi^{88}(t)=\frac{-4f_{8\eta}^{2}}{t-m_{\eta^{2}}}\ -\ \frac{4f_{8\eta'}^{2}}{t-m_{\eta'^{2}}}\ +\cdots\label{eq2.7}
\end{equation}
and a cut on the real positive t-axis running from the continuum threshold
to $\infty$.

The amplitude also possesses a QCD expansion, valid in the complex
t-plane for $|t|$ large and not too close to the physical cut. The
aim of the calculation is to relate the residues of the poles to the
QCD parameters. 
\begin{equation}
\Pi_{QCD}^{88}(t)=\Pi_{pert}^{88}\ +\ \frac{C_{1}^{88}}{t}\ +\ \frac{C_{2}^{88}}{t^{2}}\ +\cdots\label{eq2.8}
\end{equation}

The perturbative part is known to 5-loops in the chiral limit \cite{baikov2008}.

\begin{equation}
\begin{aligned}\frac{1}{\pi}Im\Pi_{pert}^{88}= & 2\frac{1}{4\pi^{2}}\{1+a_{s}+a_{s}^{2}(F_{3}+\beta_{\frac{1}{2}}L_{\mu})\\
 & +a_{s}^{3}[F_{4}+(\beta_{1}F_{3}+\frac{\beta_{2}}{2})L_{\mu}+\frac{\beta_{1}^{2}}{4}L_{\mu}^{2}]\\
 & +a_{s}^{4}[k_{3}-\frac{\pi^{2}}{4}\beta_{1}^{2}F_{3}-\frac{5}{24}\pi^{2}\beta_{1}\beta_{2}+(\frac{3}{2}\beta_{1}F_{4}+\beta_{2}F_{3}+\frac{\beta_{3}}{2})L_{\mu}\\
 & +\frac{\beta_{1}}{2}(\frac{3}{2}\beta_{1})F_{3}+\frac{5}{4}\beta_{2})L_{\mu}^{3}+\frac{\beta_{1}^{3}}{8}L_{\mu}^{3}]
\end{aligned}
\label{eq2.9}
\end{equation}
where 

$a_{s}=\frac{\alpha_{s}(\mu^{2})}{\pi},\qquad L_{\mu}=ln(\frac{-t}{\mu^{2}}),\qquad\beta_{1}=-\frac{1}{2}(11-\frac{2}{3}n_{f}),\qquad\beta_{2}=-\frac{1}{8}(102-\frac{38}{3}n_{f}),$
\\

$\beta_{3}=-\frac{1}{32}(\frac{2857}{2}-\frac{5033}{18}n_{f}+\frac{325}{54}n_{f}^{2}),\qquad F_{3}=1.9857-.1153\,n_{f},\qquad F_{4}=18.2427-\frac{\pi^{2}}{3}(\frac{\beta_{1}}{2})^{2}-4.2158\,n_{f}+.0862\,n_{f}^{2}$
\\

and $k_{3}=49.076$. \\

The strong coupling constant is likewise known to 5-loop order \cite{chetyrkin1997strong}
in terms of $\frac{\alpha_{s}^{(1)}}{\pi}\equiv\frac{-2}{\beta_{1}L}$
with $L=ln(\frac{-t}{\Lambda^{2}})$ where $\Lambda^{2}$ defines
the standard $\overline{MS}$ scale to be used here. 
\begin{equation}
C_{1}^{88}=\frac{2}{\pi^{2}}(1+2a_{s})\,m_{s}^{2}\label{eq2.10}
\end{equation}
is a correction to the perturbative part proportional to $m_{s}^{2}$
\cite{braaten1992qcd} and 
\begin{equation}
\begin{aligned}C_{2}^{88} & =\frac{1}{6}(1-\frac{11}{18}a_{s})\,\langle a_{s}G\tilde{G}\rangle+\frac{8}{3}(1-\frac{7}{3}a_{s}-\frac{75}{6}a_{s}^{2})\,\langle m_{s}\bar{s}s\rangle\\
\\
C_{3}^{88} & =\frac{-448}{\pi^{2}}a_{s}{\langle\bar{u}u\rangle}^{2}
\end{aligned}
\label{eq2.11}
\end{equation}

\noindent Consider next the contour $C$ shown in figure 1 consisting
of two straight lines parallel to the real axis and located just above
and just below the cut and running from the continuum threshold to
a large value $R$ and the circle of radius $R$.

\begin{figure}[H]
\centering{} \includegraphics[width=11cm]{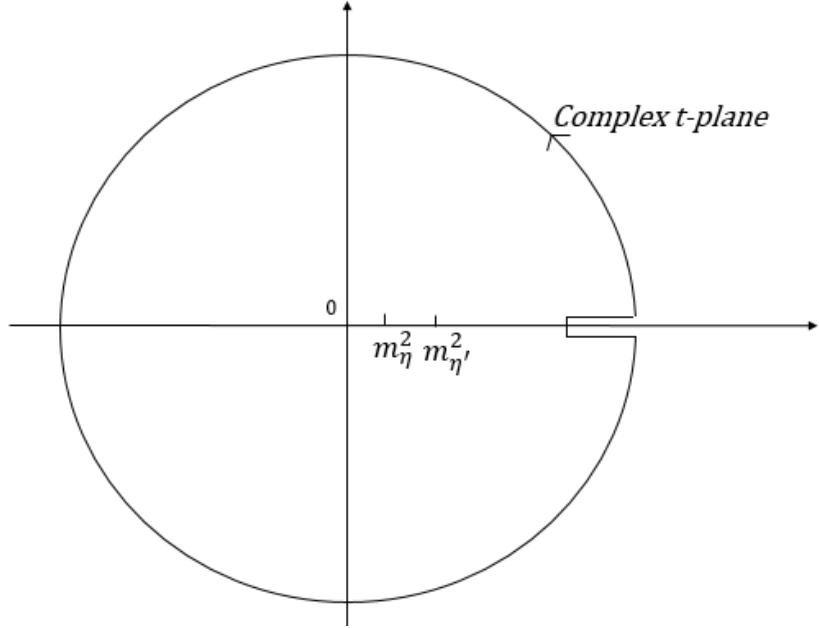} \caption{The contour of integration C.}
\label{fig1} 
\end{figure}

\noindent And consider the integral 
\[
\int_{c}dt\,f(t)\,\Pi(t)
\]
where $f(t)$ is an entire function. On the circle $\Pi(t)$ can be
replaced by $\Pi_{QCD}(t)$ to a good approximation.

Application of Cauchy's theorem leads to 
\begin{equation}
\begin{aligned}4f_{8\eta}^{2}\,f(m_{\eta}^{2})+4f_{8\eta'}^{2}\,f(m_{\eta'}^{2})= & -\frac{1}{\pi}\int_{th}^{R}dt\,f(t)\,Im\,\Pi(t)\\
 & -\frac{1}{2\pi i}\oint dt\,f(t)\,\Pi_{pert}(t)-\frac{1}{2\pi i}\oint dt\,f(t)\,\Pi_{np}(t)
\end{aligned}
\label{eq2.12}
\end{equation}

The first term on the r.h.s of the equation above, which represents
the contribution of the physical continuum constitutes the main uncertainty
of the calculation. The choice of the so-far arbitrary entire function
f(t) aims at reducing this term as much as possible in order to allow
its neglect. All that is known about the continuum is that it is dominated
by the pseudoscalar excitations $\eta(1295)$ and $\eta(1440)$ as
well as the axial-vector isoscalars $f_{1}(1285)$ and $f_{1}(1420)$
with practically the same masses.

I shall choose for a $f(t)$ simple polynomial 
\[
f(t)=p(t)=1-a_{1}t-a_{2}t^{2}
\]
the coefficients $a_{1}$ and $a_{2}$ of which annihilate $p(t)$
at the masses of the resonances, i.e 
\begin{equation}
p(t)=1-1.090GeV^{-2}t+.294GeV^{-4}t^{2}\label{eq2.13}
\end{equation}
with this choice the integrand is reduced to only a few percent of
its initial value on the interval $1.5GeV^{2}\leqslant t\leqslant2.5GeV^{2}$
and the contribution of the continuum is thus practically annihilated.

$\Pi_{pert}(t)$ has a different analytical structure than the physical
amplitude, it has a cut on the real t-axis which starts at the origin
so that $\frac{1}{2\pi i}\oint_{c'}dt\,f(t)\Pi_{pert}(t)=0$ where
$C'$ is the contour shown in figure 2

\begin{figure}[H]
\centering{} \includegraphics[width=11cm]{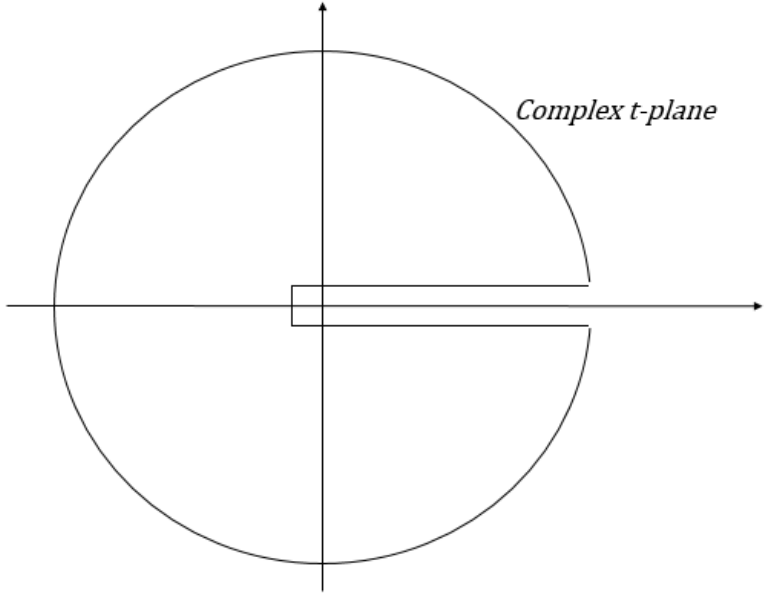} \caption{The contour of integration C' used to transform the integral $\Pi_{pert}(t)$
over the circle into an integral over the real axis.}
\label{fig2} 
\end{figure}

It then follows that 
\begin{equation}
\frac{1}{2\pi i}\oint dt\,f(t)\Pi_{pert}(t)=-\frac{1}{\pi}\int_{0}^{R}dt\,f(t)Im\,\Pi_{pert}(t)
\end{equation}
Also

\begin{equation}
\begin{aligned}\frac{1}{2\pi i}\oint dt\,f(t)\Pi_{rep}(t) & =-\frac{1}{2\pi i}\oint dt\,(1-a_{1}t-a_{2}t^{2})(\frac{C_{1}^{88}}{t}+\frac{C_{2}^{88}}{t^{2}}+\frac{C_{3}^{88}}{t^{3}}+...)\\
\\
 & =C_{1}^{88}-a_{1}C_{2}^{88}-a_{2}C_{3}^{88}
\end{aligned}
\end{equation}

The second term on the r.h.s of eq.(\ref{eq2.12}) equals the contribution
of the integral over the circle of $\Pi_{pert}(t)$ and provides the
main contribution. The last two terms are contributed by the corresponding
ones in eq.\ref{eq2.8}. Thus

\begin{equation}
4f_{8\eta}^{2}p(m_{\eta}^{2})+4f_{8\eta'}^{2}p(m_{\eta'}^{2})=\frac{1}{\pi}\int_{0}^{R}dt\quad p(t)Im\Pi_{pert}(t)-C_{1}^{88}+a_{1}C_{2}^{88}+a_{2}C_{3}^{88}\label{eq2.14}
\end{equation}
The choice of $R$ is determined by stability considerations. It should
not be too small as this would invalidate the OPE on the circle nor
should it be too large because $p(t)$ would start enhancing the contribution
of the continuum instead of suppressing it. We seek an intermediate
range of $R$ for which the integral in eq.(\ref{eq2.14}) is stable.
This turns out to be the case for $1.5GeV^{2}\leqslant R\leqslant2.5GeV^{2}$.
The integral provides the main contribution to the r.h.s of eq.(\ref{eq2.14}).\\

A similar treatment of the amplitude $\Pi^{00}(t)$ leads to 
\begin{equation}
4f_{0\eta}^{2}p(m_{\eta}^{2})+4f_{0\eta'}^{2}p(m_{\eta'}^{2})=\frac{1}{\pi}\int_{0}^{R}dt\quad p(t)Im\Pi_{pert}^{00}(t)-C_{1}^{00}+a_{1}C_{2}^{00}+a_{2}C_{3}^{00}\label{eq2.15}
\end{equation}
where $\Pi_{pert}^{00}=\Pi_{pert}^{88}$ and $C_{1}^{00}$ and $C_{2}^{00}$
are the non-perturbative coefficients of the QCD expansion 
\begin{equation}
\begin{aligned}\Pi_{QCD}^{00}(t) & =\Pi_{pert}^{00}(t)+\frac{C_{1}^{00}}{t}+\frac{C_{2}^{00}}{t^{2}}+...\\
C_{1}^{00} & =\frac{1}{\pi^{2}}(1+2a_{s})m_{s}^{2}\\
C_{2}^{00} & =\frac{1}{6}(1-\frac{11}{18}a_{s})\langle a_{s}GG\rangle+\frac{4}{3}(1-\frac{7}{3}a_{s}-\frac{75}{6}a_{s}^{2})\langle m_{s}\bar{s}s\rangle\\
C_{3}^{00} & =-\frac{448}{81}\pi^{2}a_{s}{\langle\bar{u}u\rangle}^{2}
\end{aligned}
\label{eq2.16}
\end{equation}

Finally turn to the mixed amplitude $\Pi^{08}(t)$, with the result
\begin{equation}
4f_{8\eta}f_{0\eta}p(m_{\eta}^{2})+4f_{8\eta'}f_{0\eta'}p(m_{\eta'}^{2})=-C_{1}^{08}+a_{1}C_{2}^{08}+a_{2}C_{3}^{08}\label{eq2.17}
\end{equation}
with 
\begin{equation}
\begin{aligned}C_{1}^{08}= & \frac{-\sqrt{2}}{\pi^{2}}(1+2a_{s})m_{s}^{2}\\
C_{2}^{08}= & \frac{-8\sqrt{2}}{3}(1-\frac{7}{3}a_{s}-\frac{75}{6}a_{s}^{2})\langle m_{s}\bar{s}s\rangle\\
C_{3}^{08}\simeq & 0
\end{aligned}
\label{eq2.18}
\end{equation}

Eq.(\ref{eq2.18}) is distinguished from eqs.(\ref{eq2.14}) and (\ref{eq2.15})
in that the dominant perturbative contribution is now absent and the
smallness of its r.h.s. will result in the smallness of the $\eta-\eta'$
mixing, i.e of the couplings $f_{0\eta}$ and $f_{8\eta'}$.

Eqs.(\ref{eq2.14}), (\ref{eq2.15}) and (\ref{eq2.17}) are however
insufficient to determine all four couplings. An additional equation
is obtained by considering the integral $\dfrac{1}{2\pi i}\int_{c}dt\quad tp(t)\Pi^{08}(t)$.\\

The fast convergence of the amplitude, due to the absence of th perturbative
part in the asymptotic expansion guarantees the reliability of the
result. This yields 
\begin{equation}
4f_{8\eta}f_{0\eta}p(m_{\eta}^{2})m_{\eta}^{2}\quad+\quad4f_{8\eta'}f_{0\eta'}p(m_{\eta'}^{2})m_{\eta'}^{2}\quad=\quad-C_{2}^{08}+a_{1}C_{3}^{08}\label{eq2.19}
\end{equation}

The numbers used for the condensates are

$m_{s}=(.10\pm.01)\ GeV$\\

$-\langle\bar{s}s\rangle=(.012\pm.002)\ GeV^{3}$\\

$\langle a_{s}G\tilde{G}\rangle=.013\ GeV^{4}$\\

and the value of the integral in eqs.(\ref{eq2.14}), (\ref{eq2.15})
at the stability values of $R$

$\frac{1}{\pi}\int_{0}^{R}dtp(t)Im\Pi_{pert}(t)=.034\ GeV^{2}$ as
appears in figure 3

\begin{figure}[H]
\centering{}\includegraphics[width=14cm]{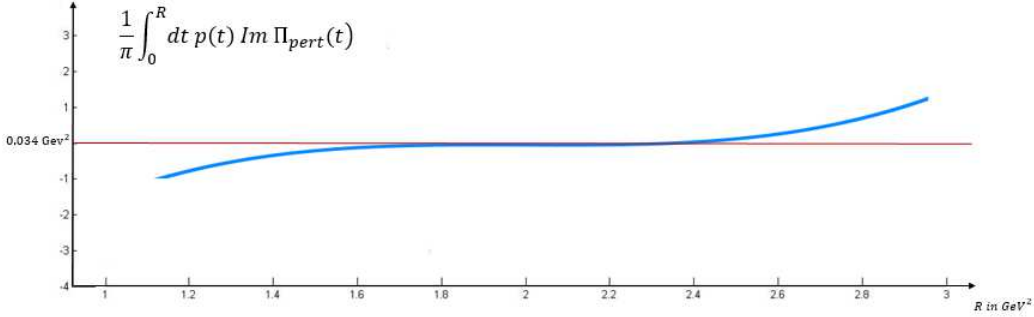} \caption{The variation of $\frac{1}{\pi}\int_{0}^{R}dt\,f(t)Im\,\Pi(t)$ as
a function of R.}
\label{fig3} 
\end{figure}

These finally yield for the couplings

\begin{equation}
\begin{aligned}f_{8\eta} & =.104\ GeV\quad\quad f_{8\eta'}= & -.046\ GeV\\
f_{0\eta} & =.042\ GeV\quad\quad f_{0\eta'}= & .160\ GeV
\end{aligned}
\label{eq:2.20}
\end{equation}
which correspond to mixing angles 
\begin{equation}
\theta_{8}=tan^{-1}(\frac{f_{8\eta'}}{f_{8\eta}})=-24^{\circ}\quad\quad and\quad\quad\theta_{0}=tan^{-1}(\frac{-f_{0\eta}}{f_{0\eta'}})=-14.7^{\circ}
\end{equation}

The values obtained above can be used in the calculation of the corrections
to the Gell-Mann-Oakes-Renner relation \cite{PhysRev.175.2195} in
the isoscalar channel. A Ward identity introduces a subtraction which
improves the convergence of the dispersion relation and therefore
their reliability.

Start with the correlator

\begin{equation}
T^{88}(t)=\int idx\quad e^{iqx}\langle0|TD^{(8)}(x)D^{(8)}(0)|0\rangle
\end{equation}
where $D^{(8)}=\partial_{\mu}A_{\mu}^{(8)}$ 
\begin{equation}
T^{88}(t)=\frac{-4f_{8\eta}^{2}m_{\eta}^{4}}{t-m_{\eta}^{4}}\quad-\quad\frac{4f_{8\eta'}^{2}m_{\eta'}^{4}}{t-m_{\eta'}^{4}}\quad+\cdots
\end{equation}
which satisfies the Ward identity

\noindent 
\[
T^{88}(0)=-\frac{16}{3}\langle m_{s}\bar{s}s\rangle
\]

Introducing a subtraction consists in considering the integral $\frac{1}{2\pi}\int_{c}\frac{dt}{t}p(t)\Pi^{88}(t)$

This gives 
\begin{equation}
\begin{aligned}f_{8\eta}^{2}m_{\eta}^{2}p(m_{\eta}^{2})+f_{8\eta'}^{2}m_{\eta'}^{2}p(m_{\eta'}^{2})= & -\frac{4}{3}\langle m_{s}\bar{s}s\rangle+m_{s}^{2}\{\frac{1}{2\pi^{2}}(1+\frac{17}{3}a_{s})\int_{0}^{R}dtp(t)\\
 & +\frac{4}{3}a_{1}(2\langle m_{s}\bar{s}s\rangle-\frac{1}{4}\langle a_{s}G\tilde{G}\rangle)\}
\end{aligned}
\end{equation}

Numerically

\[
f_{8\eta}^{2}m_{\eta}^{2}p(m_{\eta}^{2})=.002\ GeV^{4}
\]
which results in recovering $m_{\eta}$ 
\begin{equation}
m_{\eta}=(500\pm30)\ MeV
\end{equation}

The uncertainly is estimated from the one in the parameters.

\section{The topological susceptibility and its derivative at zero momentum
transfer}

The topological susceptibility

\begin{equation}
\chi(t)=i\int dx\,e^{iqx}\langle0|T\,Q(x)\,Q(0)\,|\,0\rangle
\end{equation}
has poles at the pseudoscalar mesons

\begin{equation}
\chi(t)=-\frac{{\langle0|Q|\pi\rangle}^{2}}{t-m_{\pi}^{2}}-\frac{{\langle0|Q|\eta\rangle}^{2}}{t-m_{\eta}^{2}}-\frac{{\langle0|Q|\eta'\rangle}^{2}}{t-m_{\eta}^{2}}+\cdots
\end{equation}

Consider again the integral $\frac{1}{2\pi i}\int_{c}\frac{dt}{t}p(t)\chi(t)$
with the same polynomial $p(t)$ introduces in order to suppress the
contribution of the physical continuum, it gives 
\begin{equation}
\begin{aligned}\chi(0)=\frac{{\langle0|Q|\pi^{0}\rangle}^{2}}{m_{\pi}^{2}}+\frac{{\langle0|Q|\eta\rangle}^{2}}{m_{\eta}^{2}}p(m_{\eta}^{2})+\frac{{\langle0|Q|\eta'\rangle}^{2}}{m_{\eta'}^{2}}p(m_{\eta'}^{2})+\frac{1}{2\pi i}\oint\frac{dt}{t}p(t)\,\chi^{QCD}(t)\end{aligned}
\label{eq:3.3}
\end{equation}
and for the derivative 
\begin{equation}
\begin{aligned}\chi'(0)-a_{1}\chi(0)=\frac{{\langle0|Q|\pi^{0}\rangle}^{2}}{m_{\pi}^{4}}+\frac{{\langle0|Q|\eta\rangle}^{2}}{m_{\eta}^{4}}p(m_{\eta}^{2})+\frac{{\langle0|Q|\eta'\rangle}^{2}}{m_{\eta'}^{4}}+\frac{1}{2\pi i}\oint\frac{dt}{t^{2}}p(t)\,\chi^{QCD}(t)\end{aligned}
\end{equation}

The coupling $\langle0|Q|\pi^{0}\rangle$ is given in \cite{PhysRevD.19.2188}

\begin{equation}
\langle0|Q|\pi^{0}\rangle=\frac{i}{4}f_{\pi}m_{\pi}^{2}(\frac{m_{d}-m_{u}}{m_{d}+m_{u}})
\end{equation}
and the couplings $\langle0|Q|\eta\rangle$ and $\langle0|Q|\eta'\rangle$
are obtained by sandwiching eq.(\ref{eq2.3}) between the vacuum and
the $\eta,\eta'$ states

\begin{equation}
\begin{aligned}\langle0|Q|\eta\rangle= & \sqrt{\frac{1}{12}}(f_{8\eta}+\sqrt{2}f_{0\eta})m_{\eta}^{2}\\
\langle0|Q|\eta'\rangle= & \sqrt{\frac{1}{12}}(f_{8\eta}+\sqrt{2}f_{0\eta'})m_{\eta'}^{2}
\end{aligned}
\end{equation}

The QCD expression is \cite{ioffe2000correlation,pasupathy2006derivative,singh2009axial}
\begin{equation}
\chi^{QCD}(t)=C_{21}\,t^{2}ln-t+C_{22}\,t^{2}{(ln-t)}^{2}+C_{01}\,ln-t+C_{00}+\frac{C_{-1}}{t}+\frac{C_{-2}}{t^{2}}+I(t)
\end{equation}
where $I(t)$ stands for the instanton contribution and

\begin{equation}
\begin{aligned}C_{21}= & -(\frac{\alpha_{s}}{8\pi})^{2}\frac{2}{\pi^{2}}\,(1+\frac{83}{4}\frac{\alpha_{s}}{\pi})\\
C_{22}= & \frac{9}{4}(\frac{\alpha_{s}}{\pi})\,C_{21}\\
C_{01}= & \frac{9}{64}{(\frac{\alpha_{s}}{\pi})}^{2}\,\langle\frac{\alpha_{s}}{\pi}G\tilde{G}\rangle\\
C_{-1}= & -\frac{1}{8}(\frac{\alpha_{s}}{\pi})\langle g_{s}\frac{\alpha_{s}}{\pi}G^{3}\rangle\\
C_{-2}= & -\frac{15}{128}\pi^{2}\,(\frac{\alpha_{s}}{\pi})\,{\langle\frac{\alpha_{s}}{\pi}G\tilde{G}\rangle}^{2}\\
C_{00}= & -\frac{1}{16}(\frac{\alpha_{s}}{\pi})\,\langle\frac{\alpha_{s}}{\pi}G\tilde{G}\rangle
\end{aligned}
\end{equation}
when calculations are curried out and numbers inserted eq.(\ref{eq:3.3})
yields 
\begin{equation}
\chi(0)=.94\:.10^{-3}\ GeV^{4}+\delta_{1}
\end{equation}
where 
\begin{equation}
\delta_{1}=\frac{1}{2\pi i}\oint\frac{dt}{t}\,p(t)\,I(t)
\end{equation}
denotes the instanton contribution. For the derivative 
\begin{equation}
\chi'(0)=a_{1}\chi(0)+2.31\:.10^{-3}\ GeV^{2}+\delta_{2}
\end{equation}
with 
\begin{equation}
\delta_{2}=\frac{1}{2\pi i}\oint\frac{dt}{t^{2}}\,p(t)\,I(t)
\end{equation}

The instanton term $I(t)$ is model dependent, the form used by Ioffe
and Samsonov \cite{ioffe2000correlation} is 
\[
I(t)=t^{2}\int dp\,n(\rho)\,\rho^{4}\,K_{2}^{2}\,(Q\rho)
\]
where 
\begin{equation}
n(\rho)=n_{0}\,\delta(\rho-\rho_{c}),\qquad\rho_{c}=1.5\ GeV^{-1}\label{eq3.13}
\end{equation}
and $K_{2}(Q\rho)$ is the MacDonald function. It should be noted
however that important screening corrections, as has been emphasized
by Forkel \cite{forkel2005direct}, can modify considerably expression
eq.(\ref{eq3.13}).

I shall take the screening corrections into account simply by considering
the overall factor as a free parameter to be determined by the calculation.
Thus let

\begin{equation}
I(t)=c\,t^{2}\,K_{2}^{2}\,(\rho_{c}\sqrt{-t})
\end{equation}

In order to proceed further the constant $c$ has to be determined.
This is done by considering the integral $\frac{1}{2\pi i}\int_{c}dtp(t)\chi(t)$
because the only poles of the integrand lie at the pseudoscalars we
have

\begin{equation}
0=\langle0\left\vert Q\right\vert \pi\rangle^{2}+\langle0\left\vert Q\right\vert \eta\rangle^{2}p(m_{\eta}^{2})+\langle0\left\vert Q\right\vert \eta^{\prime}\rangle^{2}p(m_{\eta^{\prime}}^{2})+\delta_{0}
\end{equation}
with
\begin{equation}
\delta_{0}=\frac{1}{2\pi i}\int_{c}dtp(t)I(t)=\frac{c}{2\pi i}\int_{c}dtp(t)t^{2}K_{2}^{2}(\rho_{c}\sqrt{-t})
\end{equation}

Asymptotic forms of $K_{2}(x)$ are given in Dwight \cite{dwight1947tables}
these are used to evaluate the integral above which yields

\begin{equation}
c=-.376\,.10^{-3}
\end{equation}

This together with a similar evaluation of the corresponding integrals
appearing in the expressions of $\delta_{1}$ and $\delta_{2}$ give

\begin{equation}
\delta_{1}=.177\,.10^{-3}GeV^{4}\quad and\quad\delta_{2}=.028\,.10^{-3}GeV^{6}
\end{equation}
which corresponds to

\begin{equation}
\chi(0)=1.10.10^{-3}GeV^{4}\quad and\quad\chi^{\prime}(0)=3.5.10^{-3}GeV^{2}
\end{equation}

The value obtained for $\chi(0)$ is quite close to the one computed
on the lattice \cite{PhysRevLett.94.032003} $\chi(0)=1.33.10^{-3}GeV^{4}$
and to the one given by the Witten-Veneziano \cite{veneziano1979u,witten1979current}
formula obtained in the large $N_{c}$ limit
\begin{equation}
\chi(0)=\frac{f_{\pi}^{22}}{2n_{f}}(m_{\eta}^{2}+m_{\eta^{\prime}}^{2}-2m_{K}^{2})=1.05.10^{-3}GeV^{4}
\end{equation}

As to $\chi^{\prime}(0)$ the value obtained is relatively large,
close to the one advocated by Ioffe \cite{ioffe2000correlation},
$\chi^{\prime}(0)=(2.9$\ $\pm.4).10^{-3}GeV^{2}$

\section{Results and Conclusion}

The subject of octet-singlet mixing of the pseudoscalar mesons has
been studied and the couplings of the $\eta$ and $\eta'$ mesons
to the axial-currents $A_{\mu}^{0}$ and $A_{\mu}^{8}$ evaluated
yielding for the mixing angles $\theta_{8}=-24^{\circ}$ and $\theta_{0}=-14.7^{\circ}$.
The corrected GMOR relation reproduces the value of $m_{\eta}$. The
topological susceptibility and its derivative at the origin have also
been computed with the effects of instantons and instanton screening
taken into account resulting in $\chi^{\prime}(0)=3.5\,.10^{-3}GeV^{2}$
, $\chi^{\prime}(0)=1.05\,.10^{-3}GeV^{2}$\newpage{}

\end{document}